\documentclass[11pt,twoside]{amsart}
\usepackage{amsmath, amsthm, amscd, amsfonts, amssymb, graphicx, color}
\usepackage[bookmarksnumbered, colorlinks, plainpages]{hyperref}

\newtheorem{thm}{Theorem}[section]
\newtheorem{cor}[thm]{Corollary}
\newtheorem{lem}[thm]{Lemma}

\newtheorem{defn}[thm]{Definition}

\numberwithin{equation}{section}
\def\pn{\par\noindent}

\begin{document}

\leftline{ \scriptsize \it Bulletin of the Iranian Mathematical
Society  Vol. {\bf\rm XX} No. X {\rm(}201X{\rm)}, pp XX-XX.}

\vspace{1.3 cm}

\title{A new proof for the Banach-Zarecki theorem: A light on
integrability and continuity}
\author{A. Mahdipour--Shirayeh$^*$ and H. Eshraghi}

\thanks{{\scriptsize
\hskip -0.4 true cm MSC(2010): Primary: 26B30; Secondary: 46B22, 46G12.
\newline Keywords: Banach--Zarecki theorem, Radon--Nikodym
theorem, Lusin's condition.\\
Received: 20 September 2011, Accepted: 3 June 2012.\\
$*$Corresponding author
\newline\indent{\scriptsize $\copyright$ 2012 Iranian Mathematical
Society}}}

\maketitle

\begin{center}
Communicated by\;
\end{center}

\begin{abstract}
To demonstrate more visibly the close relation between the
continuity and integrability, a new proof for the Banach-Zarecki
theorem is presented on the basis of the Radon-Nikodym theorem
which emphasizes on measure-type properties of the Lebesgue
integral. The Banach-Zarecki theorem says that a real-valued
function $F$ is absolutely continuous on a finite closed interval
if and only if it is continuous and of bounded variation when it
satisfies Lusin's condition. In the present proof indeed a more
general result is obtained for the Jordan decomposition of $F$.
\end{abstract}

\vskip 0.2 true cm


\pagestyle{myheadings}
\markboth{\rightline {\scriptsize  Mahdipour--Shirayeh and Eshraghi}}
         {\leftline{\scriptsize A new proof for the Banach-Zarecki theorem}}

\bigskip
\bigskip


\section{\bf Introduction}
\vskip 0.4 true cm

The original motivation for the present work concerns with the
open debate of the regularity of hydrodynamical parameters of
fluid flows. It is still not known that starting from a smooth
initial conditions in a three dimensional fluid, when and how any
kind of blow up or singularity will happen. A large amount of
works consider this problem in various special cases and obtain
many results. It was known that the type of singularity is so
strong such that many kinds of integral norms of hydrodynamical
quantities are also singular. However, almost all of these
integral norms are obtained by the Lebesgue integration but we
know that there are other types of integration that are
generalizations of the usual Riemann integral and do not coincide
the Lebesgue integral.

So, a natural question comes that how can
we say something when our functions are not Lebesgue integrable?
How should one replace (absolute) continuities and regularities in
these new cases? As the first step it looks necessary to test and
generalize a direct relation between the integration and
continuity and Banach--Zarecki Theorem provides perhaps the most
visible case to observe such a relation. It was therefore needed
to discover a more direct and closer relation between the absolute
continuity and the Lebesgue integral to be an arrow for other
works.

Banach--Zarecki Theorem is a classical theorem in real analysis
with many applications mostly in geometric and functional analysis
as well as some physical and engineering subjects. The origin of
this theorem was stated and proved by Banach and independently by
Zarecki for a real--valued function on an interval \cite{N}. For
functions of a real variable with values in reflexive Banach
spaces, the result is contained in \cite{F}, Theorem 2.10.13,
where the codomain space has the Radon-Nikodym property. There
also exists another version of the theorem initiated by an old
result of Lusin \cite{L}, later extended for a function of a real
variable with values in a metric space \cite{Du,DZ}.

It is not surprising that there is a variety of extensions for this theorem
to more variables in many ways and also by natural changes in
properties well-known in one dimensional case such as almost
everywhere continuity and differentiability, integration by parts
and so on \cite{DZa,Ji,Ma}. In fact this theorem can be
generalized to the concept of approximate continuity that plays an
important role to understand the relationship between Riemann
integrability (for almost everywhere continuous functions) and
continuity on the one hand, and the relationship between
approximate continuity and Lebesgue integrability (for almost
everywhere approximately continuous functions), on the other hand
\cite{DZ}.

There exist alternative proofs for this theorem; although these
are of different appearance but they are constructed from a common
root (see e.g. \cite{Bru,Car,Roy,Yeh}). In the present work the
classical form of the theorem is considered, since it looks
possible to naturally extend the results to more general cases
mentioned above. The most convenient statement of the
Banach--Zarecki theorem is \cite{Bru}:
\begin{thm}{\label{th1}} {\it Let $F$ is a real--valued function defined on
a real bounded closed interval $[a, b]$. A necessary and
sufficient condition for $F$ to be absolutely continuous is that\\
$(i)$ $F$ is continuous and of bounded variation on $[a, b]$,\\
$(ii)$ $F$ satisfies Lusin's condition, i.e. it maps sets of
Lebesgue measure zero into sets of Lebesgue measure zero.}
\end{thm}
The necessary condition is straightforward and will not be
discussed here. Its proof is given in almost any text book of real
analysis \cite{Bru,JZ}. However the sufficient condition is rather
technical and requires some non--trivial efforts and may rarely be
found in common references. Thus, our attempt is concentrated on
providing an alternative proof of the sufficient condition, that
is, if a real--valued function is continuous and of bounded
variation and also satisfies Lusin's condition, then it is
absolutely continuous. In \cite{Bru}, there is a proof for the
sufficient condition employing an inequality being also proved in
this reference. The main tools of this approach are the almost
everywhere differentiability and the Vitali covering theorem.

However the present proof is based on the close relation between
the Lebesgue integral and the properties of a measure space which
manifests itself essentially through the Radon-Nikodym theorem.
Thus, the main used tools here are the Radon-Nikodym theorem and
the properties of variations of functions. This new proof may
however cost to be considered because of several reasons such as
the following. Here a slightly more general result is proven,
namely Lemma \ref{4} while we need only Corollary \ref{5} for our
proof. The concept of almost everywhere differentiability and thus
the Vitali covering lemma is not used. The methods and techniques
handled here seem to be applicable and naturally generalizable to
a class of similar problems. There is a hope to generalize this
method to obtain an analog version for the absolute continuity in
relation with other types of integration rather than the Lebesgue
integral.

Finally it is seen that here some statements are proven
employing only conditions $(i)$ and $(ii)$ mentioned in the
Banach--Zarecki theorem and without using the absolute continuity
condition, while these statements are usually proved through a
direct application of the absolute continuity condition in the
common literatures.

In order to prove Theorem \ref{th1}, our strategy is to establish
the following theorem which illustrates more clearly, the relation
between the absolute continuity and the Lebesgue integral.
\begin{thm}\label{main}  Suppose that $F:[a,b]\longrightarrow{\Bbb R}$
is a continuous and of bounded variation and satisfies Lusin's
condition. Then there exists an integrable function and in fact a
Borel--measurable function $f:[a,b]\longrightarrow{\Bbb R}$ such
that
\begin{align*}
F(x)=F(a)+\int_{[a,x]}\,f\,d\lambda~~:~~~~\forall x\in [a,b],
\end{align*}
where $d\lambda$ in the integral comes from the Lebesgue measure $\lambda$.
\end{thm}
This theorem will immediately yield Theorem \ref{th1} through the
application of the well known statement \cite{Bru,JZ}:
\begin{quote}
{\em Let $f:[a,b]\longrightarrow{\Bbb R}$ be a Lebesgue integrable
function and let $F(x)=F(a) + \int_{[a,x]}f\,d\lambda$, then $F$
is absolute continuous on $[a,b]$.}
\end{quote}

In the next section, we prove the Theorem \ref{main} in
three steps, the first of which is well known in text books
\cite{JZ} while step 2 and especially step 3 are of our main
interests.

Throughout this paper we assume that the notation $\lambda$
implies the Lebesgue measure, unless specially stated otherwise.


\section{\bf The main result:
new proof of Theorem 1.2}
\vskip 0.4 true cm

The proof is divided into three interconnected steps.\\

{\em Step 1.} At first, we prove the theorem assuming that $F$ is
strictly increasing. In this case, the proof coincides the
standard proof given in common text books (see e.g. Theorem 4.3.8
of \cite{JZ}) which employs the Radon--Nikodym theorem. To have a
complete discussion, let us briefly review the proof here.

Since $F$ is strictly increasing, $F$ is a homeomorphism from $I=[a,b]$
to $J=F(I)=[F(a),F(b)]$ and so $F$ preserves Borel sets between
$I$ and $J$. Let ${\mathcal B}$ be the collection of Borel measurable
subsets of $I$, then we can define the new measure $\nu:{\mathcal
B}\longrightarrow [0,\infty)$ as $\nu(E)=\lambda(F(E))$. It is
clear that $\nu$ is a finite measure and is absolutely continuous
relative to $\lambda$ (since $F$ satisfies Lusin's condition).
Therefore, according to the Radon--Nikodym theorem, there exists a
(Borel) measurable and Lebesgue integrable function
$f:I\longrightarrow {\Bbb R}$ such that
\begin{eqnarray}\label{eq:5}
\nu(E)=\int_E\,f\,d\lambda, \hspace{0.5cm} E\in {\mathcal B}.
\end{eqnarray}

Especially if $E=[a,x]$ for $x\in I$, then $F(E)=[F(a),F(x)]$ and
Eq.~\eqref{eq:5} immediately implies that
\begin{align*}
F(x)=F(a)+\int_{[a,x]}\,f\,d\lambda, \hspace{0.5cm} x\in I.
\end{align*}

This completes the proof of this step.\\

{\em Step 2.} Let $F$ is non--decreasing (i.e. increasing but not
strictly increasing). So, there exists the continuous and of
bounded variation function $G(x)=F(x)+x$ which is strictly
increasing. The proof will be complete if we prove that Lusin's
property is fulfilled by $G$, i.e. for $N\subset[a,b]$ if
$\lambda(N)=0$ then $\lambda(G(N))=0$. Since $F$ is
non--decreasing, one easily observes that the constant values of
$F$ make sense in disjoint intervals $S_k$ and the continuity of
$F$ implies that $S_k$s are closed intervals, say $[a_k,b_k]$.
Hence, in general, on $S=\bigcup_{k=1}^{+\infty} S_k$, $F$ takes
the values $F(S)=\Big\{\mu_k\Big\}_{k=1}^{+\infty}$ where $\mu_k$
is the value of $F$ on $S_k$.

The intervals $S_k$ may be so small and their union $S$ is not
necessary closed. Now, since $S_k$~s are disjoint, we can write
\begin{align*}
N_1=N\cap S, \hspace{1cm} N_2=N-N_1.
\end{align*}

Therefore, we have
\begin{align*}
\lambda(G(N))\leq \lambda(G(N_1))+\lambda(G(N_2)),
\end{align*}
while
\begin{align*}
\lambda(G(N_1))= \lambda\Big( \bigcup_{k=1}^{+\infty}G(N\cap S_k)
\Big) \leq \sum_{k=1}^{+\infty}\lambda (G(N\cap S_k)).
\end{align*}

On the other hand, $G(N\cap S_k)=\{ \mu_k + x~|~x\in N\cap S_k\}$
and thus $\lambda(G(N\cap S_k))=\lambda(N\cap S_k)$, so
\begin{align*}
\lambda(G(N_1)) &\leq \sum_{k=1}^{+\infty}\lambda (G(N\cap
S_k)) \nonumber\\
&= \lambda(\bigcup_{k=1}^{+\infty}(N\cap S_k))=\lambda(N_1)\leq
\lambda(N)=0.
\end{align*}

Therefore $\lambda(G(N_1))=0$. To prove $\lambda(G(N_2))=0$, we
notice that $F$ satisfies Lusin's condition i.e. $\lambda(N_2)=0$
results in $\lambda(F(N_2))$, so for each $\epsilon>0$, we can
find an open set $U$ such that $F(N_2)\subset U$ with
$\lambda(U)<\epsilon$. In addition, since $\lambda(N_2)=0$, one
can find an open set $U'$ including $N_2$ such that
$\lambda(U')<\epsilon$. The open set $V:= U'\cap F^{-1}(U)$
contains $N_2$ such that $\lambda(V)<\epsilon$ and
$\lambda(F(V))<\epsilon$. Suppose $V=\bigcup_{k=1}^{+\infty}I_k$
where $I_k$s are disjoint open intervals. For each $I_k$, consider
the two closed intervals (if exist) $S_i$ and $S_j$ intersecting
$I_k$ from the left and right containing the left and right
boundary points of $I_k$ resp. Define $I'_k = I_k-(S_i\cup S_j)$
(it is possible that $I'_k$ is empty). Thus $I'_k\subset I_k$ and
$I'_k$~s are mutually disjoint.

Let $V':=\bigcup_{k=1}^{+\infty}I'_k$ and since $S_k$~s are all
out of $N_2$, $N_2\subset V'$ and thus $F(N_2)\subset F(V')$.
It is important to attend that for each $l$ and $k$, $S_l$ is
either completely contained in $I'_k$ or is disjoint from it.
According to the conditions on $F$, i.e. non--increasing and
continuity, one can deduce that $F(I'_k)$ is an interval (not
necessarily closed or open) which we denote it by $J_k$. Now we
acclaim that $J_k$~s are mutually disjoint. If else, for example
if $y\in J_k\cap J_l$ for some $k$ and $l$, then there exist at
least two points $x_k\in I'_k$ and $x_l\in I'_l$ such that
$F(x_k)=F(x_l)$. Hence there exists an $S_i$ so that
$[x_k,x_l]\subset S_i$ but hence $S_i$ is not completely in $I'_k$
or $I'_l$ which is a contradiction. Relations
\begin{align*}
F(N_2)\subset F(V')=\bigcup_{k=1}^{+\infty}F(I'_k)\subset U,
\end{align*}
imply that
\begin{align*}
\lambda\Big(\bigcup_{k=1}^{+\infty}J_k\big)\leq \lambda(U)<
\epsilon,
\end{align*}
and since $J_k$~s are disjoint sets,
\begin{align*}
\sum_{k=1}^{+\infty}\lambda(J_k)< \epsilon.
\end{align*}

The remaining work is to determine $G(I'_k)$~s and approximate
their measure. For each $k$, we have
\begin{align*}
G(I'_k)&=\Big\{F(x)+x~|~x\in I'_k \Big\}\\ \nonumber &\subseteq
\Big\{y+x~|~y\in J_k,~x\in I'_k \Big\}\\  &\subseteq
\Big(\inf(I'_k)+\inf(J_k)~~,~\sup(I'_k)+\sup(J_k)\Big),
\end{align*}
which results in
\begin{align*}
\lambda(G(I'_k))\leq \lambda(I'_k)+\lambda(J_k).
\end{align*}

So we obtain
\begin{align*}
\lambda(G(N_2)) \leq \lambda\Big(
\bigcup_{k=1}^{+\infty}G(I'_k)\Big)\leq
\sum_{k=1}^{+\infty}\lambda(G(I'_k)).
\end{align*}

The latter equations clarify that
\begin{align*}
\lambda(G(N_2))\leq \sum_{k=1}^{+\infty}\lambda(I'_k) +
\sum_{k=1}^{+\infty}\lambda(J_k)< \epsilon + \lambda(U) <
2\epsilon.
\end{align*}

Thus $\lambda(G(N_2))=0$. This shows that Lusin's condition is
fulfilled for $G(x)=F(x)+x$. Now pertaining to Step 1, there is an
integrable and Borel-measurable function
$f_1:[a,b]\longrightarrow{\Bbb R}$ s.t.
\begin{align*}
G(x)-G(a)=\int_{[a,x]}f_1\,d\lambda,
\end{align*}
hence
\begin{align*}
F(x)-F(a)=\int_{[a,x]}f\,d\lambda,
\end{align*}
thus if we let $f=f_1-1$ and this completes the proof of Step 2.\\

{\em Step 3.} Finally, we assume that $F$ is continuous and of
bounded variation which satisfies Lusin's condition and show that
the theorem holds. To accomplish this, we make use of the
following two lemmas.
\begin{lem}{\label{3}} Let $F:[a,b]\longrightarrow{\Bbb R}$ be a continuous
function of bounded variation. If $F=p-n$ is the Jordan
decomposition for $F$, then $p$ and $n$ are continuous.
\end{lem}
\begin{lem}{\label{4}}
In the situation of Lemma \ref{3}, let $N\subset [a,b]$ such that
$F(N)$ has a zero Lebesgue measure. Then $p(N)$ and $n(N)$ are
also of Lebesgue measure zero.
\end{lem}
Lemma \ref{4} immediately yields the following result.
\begin{cor}{\label{5}}
In the situation of Lemma \ref{3}, let $F$ satisfies Lusin's
condition, then $p$ and $n$ also satisfy Lusin's condition.
\end{cor}
It is seen from Lemma \ref{3} and Corollary \ref{5} that both of
$p$ and $n$ are continuous and of bounded variation and Lusin's
condition is valid for them. Then since they are non--decreasing,
there will exist integrable and Borel--measurable real--valued
functions $g$ and $h$ on $[a,b]$ so that
$p(x)=p(a)+\int_{[a,x]}g\,d\lambda$ and
$n(x)=n(a)+\int_{[a,x]}h\,d\lambda$ and therefore the proof will
be completed substituting $f=g-h$.
%
\section{\bf Proof of Lemma 2.1}
\vskip 0.4 true cm

It is sufficient to prove that $p$ is continuous. At first, we denote that $p$ is a
right--continuous function. The continuity of $p$ can be directly
achieved by ($\epsilon-\delta$) method. However an alternative
proof is presented here because of its easier application in the
proof of Lemma \ref{4}.

According to definition,
\begin{eqnarray}\label{eq:6}
p(x)= \bigvee^x_a(F)=\sup_P |F(P)|= \sup_P \sum_{k=1}^{n(P)}
|F(x_k)- F(x_{k-1})|
\end{eqnarray}
is the variation of $F$ from $a$ to $x$ where the supremum is
taken over all partitions $$P \,:\, {a = x_0 < x_1 <  \cdots  <
x_n = x}$$ of $[a,x]$ and $n = n(P) = \#P - 1$. Therefore for
arbitrary $\epsilon > 0$ there is a partition $P$ such that
\begin{eqnarray}\label{eq:7}
0\leq \bigvee^x_a(F)- |F(P)|<\epsilon.
\end{eqnarray}
\begin{defn} For the given partition $P \,:\, a = x_0 < x_1 <  \cdots  <
x_n = b$, let $x\in [x_{i-1}, x_i]$. Two adjacent partitions
$P_1(x)$ and $P_2(x)$ are defined as
\begin{align*}
&&P_1(x) \,:\, {a=x_0 <  \cdots  < x_{i-1} \leq x},\\
&&P_2(x) \,:\, {x \leq x_i <  \cdots  < x_n=b},
\end{align*}
and partition $P'(x)$ considered as a refinement of $P$ is
\begin{align*}
P'(x):~a=x_0<\cdots<x_{i-1} \leq x \leq x_i<\cdots<x_n=b.
\end{align*}
\end{defn}
For $\epsilon>0$ and its corresponding partition $P$ considered in
Eq.~\eqref{eq:7}, one can define continuous functions ${\rm
w}_i:[x_{i-1},x_i]\longrightarrow {\Bbb R}$ as
\begin{eqnarray}\label{eq:8}
{\rm w}_i(x)= |F(P_1(x))|.
\end{eqnarray}

Application of the pasting lemma implies the existence of the
continuous function ${\rm u}_{\epsilon}: [a,b]\longrightarrow
{\Bbb R}$ so that on each $[x_{i-1},x_i]$, ${\rm u}_{\epsilon}$ is
equal to ${\rm w}_i$. Therefore
\begin{align*}
\Big(\bigvee^x_a(F)- |F(P_1(x))|\;\Big)+\Big(\bigvee^b_x(F)-
|F(P_2(x))|\;\Big)&=\bigvee^b_a(F)-|F(P'(x))|\\
&< \epsilon.
\end{align*}

The two terms on the left hand side of the above relation are
nonnegative and especially considering the first term, one finds
that
\begin{eqnarray}\label{eq:9}
0\leq p(x)- {\rm u}_{\epsilon}(x)<\epsilon,
\end{eqnarray}
in which Eqs.~\eqref{eq:7} and \eqref{eq:8} were applied.

Now consider $\Big\{{\rm u}_{_{2^{-k}}}\Big\}_{k=1}^{\infty}$ as a
sequence of continuous functions. Equation \eqref{eq:9} with
$\epsilon=2^{-k}$ shows that this sequence converges uniformly to
$p$, thus $p$ is continuous.


\section{\bf Proof of Lemma 2.2}
\vskip 0.4 true cm

Let $N\subset [a,b]$ such that $\lambda(F(N))=0$. For arbitrary $\epsilon>0$, consider its
corresponding partition $P$ as introduced in Eq.~\eqref{eq:7}. It
is sufficient to prove that $\lambda(p(N_i))=\lambda(n(N_i))=0$
where $N_i=N\cap [x_{i-1},x_i]$ ($1\leq i\leq n$). Since $F(N_i)$
has zero Lebesgue measure, there exists a sequence of disjoint
open intervals $\{J_k\}_{k=1}^{\infty}$ such that $F(N_i)\subset
\bigcup_{k=1}^{\infty} J_k$ and
\begin{eqnarray}\label{eq:10}
\sum_{k=1}^{\infty} \lambda(J_k)<\epsilon.
\end{eqnarray}

At most one of $J_k$~s contains the point $F(x_{i-1})$ and at most
one of them contains $F(x_i)$. If so, we exclude these two points
from $J_k$~s and split the interval(s) containing the points into
two adjacent open intervals. This process clearly leaves relation
\eqref{eq:10} unchanged. For each $J_k$ we have
$F^{-1}(J_k)=\bigcup_{l=1}^{\infty} I_{kl}$ where intervals
$I_{kl}=(a_{kl}, b_{kl})$ are disjoint. According to our
hypothesis, one can easily observe that
\begin{eqnarray}\label{eq:11}
\lambda(p(N_i))\leq \sum_{k,l=1}^{\infty} \lambda(p(I_{kl})).
\end{eqnarray}

Choose any finite number of intervals $I_{kl}$s and call them
$(a_1, b_1)$, $\cdots$, $(a_m, b_m)$ in such an order that we have the
partition
\begin{eqnarray}\label{eq:12}
&\hspace{1cm} Q:~b_0=x_{i-1}\leq a_1<b_1<a_2<\cdots<a_m<b_m\leq a_{m+1}=x_i.
\end{eqnarray}

Thus
\begin{align*}
\bigvee^{x_i}_{x_{i-1}}(F)-|F(Q)|<\epsilon,
\end{align*}
which means that
\begin{align*}
\sum^m_{j=1}\!\!\Big(\!\bigvee^{b_j}_{a_j}(F)-|F(b_j)-F(a_j)|\!\Big)
+\sum^m_{j=0}\!\!\Big(\!\bigvee^{a_{j+1}}_{b_j}(F)-|F(a_{j+1})-F(b_j)|\!\Big)<\epsilon.
\end{align*}

Each term in the left side is nonnegative, especially noting to
the first term and recalling the definition of $p$ through
Eq.~\eqref{eq:6} and its non-decreasing property, one concludes
that
\begin{align*}
\sum^m_{j=1} \lambda(p(a_j,b_j))<\epsilon + \sum^m_{j=1}
|F(b_j)-F(a_j)|.
\end{align*}

The above inequality holds for any finite number of $I_{kl}$~s,
thus
\begin{eqnarray}\label{eq:13}
\sum^{\infty}_{k,l=1} \lambda(p(I_{kl}))<\epsilon +
\sum^{\infty}_{k,l=1} |F(b_{kl})-F(a_{kl})|.
\end{eqnarray}

Our next task is to find an upper bound proportional to $\epsilon$
for the second term of the last equation. To do this we consider
two separate cases. The first case is when $F(x_{i-1})=F(x_i)$.
Choose again the finite number of $I_{kl}$~s say $(a_j, b_j)$~s
for $1\leq j\leq m$ and construct the partition $Q$ as introduced
in Eq.~\eqref{eq:12}. This partition is a refinement of
$x_{i-1}<x_i$ and so $|F(Q)|-|F(x_i)-F(x_{i-1})|<\epsilon$ and
thus
\begin{align*}
\sum^{m}_{j=1} |F(b_j)-F(a_j)|\leq |F(Q)|<\epsilon.
\end{align*}

The last inequality holds for any finite number of $I_{kl}$~s and
so is also valid for all of them. Therefore when
$F(x_{i-1})=F(x_i)$ by the use of Eq.~\eqref{eq:13} we have
\begin{eqnarray}\label{eq:14}
\lambda(p(N_i))\leq \sum^{\infty}_{k,l=1}
\lambda(p(I_{kl}))<2\,\epsilon.
\end{eqnarray}

The second case is related to the condition $F(x_{i-1})<F(x_i)$
(the opposite case is similar). Recall that $J_k$~s where disjoint
open intervals containing $F(N_i)$ (except probably the two points
$F(x_{i-1})$ and $F(x_i)$) with total measure less than
$\epsilon$. Thus we are able to divide them into three types:
$J_k^+$~s whose points are greater than $F(x_i)$, $J_k^-$~s whose
points are less than $F(x_{i-1})$ and $J_k^{\circ}$~s whose points
are between $F(x_{i-1})$ and $F(x_i)$. At first attend to
$J_k^+$~s. In this case, take any finite number of $I_{kl}^+$~s
(whose images are inside $J_k^+$~s) say $(a^+_j, b^+_j)$~s for
$1\leq j\leq m$ such that $x_{i-1}\leq
a^+_1<b^+_1<a^+_2<\cdots<a^+_m<b^+_m\leq x_i$. Images of these
intervals lie inside a finite (say $s$) number of $J_k^+$~s,
namely $J^+_{k_r}=(c^+_r, d^+_r)$ for $1\leq r\leq s$ where
obviously $s\leq m$.

Suppose that in addition, $(c^+_r, d^+_r)$~s
are arranged increasingly such that $F(x_i)\leq c^+_1<d^+_1
<c^+_2<\cdots<c^+_s<d^+_s$. The compact set $[x_{i-1},x_i]\cap
F^{-1}(c^+_1)$ has a minimum and maximum respectively $\alpha^+$
and $\beta^+$. Since the images of all $(a^+_j, b^+_j)$~s are
greater than $c^+_1\geq F(x_i)$, the intermediate value theorem
implies that they all lie between $\alpha^+$ and $\beta^+$. Thus
there exist partition $R_1:~x_{i-1}<\alpha^+ < \beta^+ <x_i$ and
its refinement $R_2:~x_{i-1}<\alpha^+ <a_1^+ < b_1^+ < \cdots
a_m^+ < b_m^+ < \beta^+ <x_i$. The relation
$|F(R_2)|-|F(R_1)|<\epsilon$ regarding the fact that
$F(\alpha^+)=F(\beta^+)=c^+_1$ implies that
\begin{align*}
\sum_{j=1}^m |F(b^+_j)-F(a^+_j)|<\epsilon,
\end{align*}
but since this is true for any finite number of considered
intervals, so for $I_{kl}^+=(a_{kl}^+ , b_{kl}^+)$~s we have
\begin{eqnarray}\label{eq:15}
\sum_{k,l} |F(b_{kl}^+)-F(a_{kl}^+)|<\epsilon.
\end{eqnarray}

Quite similarly, for $I_{kl}^-=(a_{kl}^- , b_{kl}^-)$~s we have
\begin{eqnarray*}\label{eq:16}
\sum_{k,l} |F(b_{kl}^-)-F(a_{kl}^-)|<\epsilon.
\end{eqnarray*}

Finally consider $J_k^{\circ}$~s whose points are between
$F(x_{i-1})$ and $F(x_i)$ where for each $k$,
$F^{-1}(J_k^{\circ})=\bigcup_{l=1}^{\infty} I^{\circ}_{kl}$.
Similar to the previous case choose a finite number of
$I^{\circ}_{kl}$~s such as $(a^{\circ}_j, b^{\circ}_j)$~s for
$1\leq j\leq m$ such that $x_{i-1}\leq
a^{\circ}_1<b^{\circ}_1<a^{\circ}_2<\cdots<
a^{\circ}_m<b^{\circ}_m\leq x_i$ and assume their images lie in
$J^{\circ}_{k_r}=(c^{\circ}_r, d^{\circ}_r)$ for $1\leq r\leq s$
where clearly $s\leq m$. Again suppose $(c^{\circ}_r,
d^{\circ}_r)$~s are arranged increasingly such that
\begin{eqnarray}\label{eq:17}
F(x_{i-1})\leq c^{\circ}_1<d^{\circ}_1<c^{\circ}_2
<\cdots<c^{\circ}_s<d^{\circ}_s\leq F(x_i).
\end{eqnarray}

Now define $\alpha^{\circ}_r=\min\Big( [x_{i-1},x_i]\cap
F^{-1}(c^{\circ}_r)\Big)$ for $1\leq r\leq s$. Relation
\eqref{eq:17} and the intermediate value theorem establish that
\begin{eqnarray}\label{eq:18}
x_{i-1}\leq \alpha^{\circ}_1<\alpha^{\circ}_2
<\cdots<\alpha^{\circ}_s<\alpha^{\circ}_{s+1}=x_i.
\end{eqnarray}

Note that in the above relation $\alpha^{\circ}_{s+1}$ is defined
to be $x_i$. In addition, define
$\beta^{\circ}_r=\max\Big( [x_{i-1},
\alpha^{\circ}_{r+1}]\\
\cap F^{-1}(d^{\circ}_r)\Big)$ for $1\leq r\leq s$ and also define
$\beta^{\circ}_0=x_{i-1}$. This definition immediately yields that
for each $r=1,\cdots s-1$ we have
$\alpha^{\circ}_r<\beta^{\circ}_r<\alpha^{\circ}_{r+1}$ while for
$r=0$ we have $x_{i-1}=\beta^{\circ}_0\leq \alpha^{\circ}_1$ and
for $r=s$ we have $\alpha^{\circ}_s<\beta^{\circ}_s\leq
\alpha^{\circ}_{s+1}=x_i$. Thus, relation \eqref{eq:18} is finally
improved to admit to define the partition
\begin{eqnarray}\label{eq:19}
&\hspace{1cm} S_1:~x_{i-1}=\beta^{\circ}_0\leq \alpha^{\circ}_1<\beta^{\circ}_1<
\alpha^{\circ}_2<\cdots<\alpha^{\circ}_s<\beta^{\circ}_s\leq
\alpha^{\circ}_{s+1}=x_i.
\end{eqnarray}

In this position we claim that for each $j, r$ ($1\leq j\leq m, 0\leq r\leq s$)
we have $(a^{\circ}_j, b^{\circ}_j)\cap(\beta^{\circ}_r,\alpha^{\circ}_{r+1})=\varnothing$.
If not, assume $y$ belongs to this set, then only two states may
happen:

In the first state we have $F(y)<d^{\circ}_r<c^{\circ}_{r+1}$ for
$1\leq r\leq s-1$, $F(y)<c^{\circ}_1$ for $r=0$ and
$F(y)<d^{\circ}_s$ for $r=s$. The case $r=0$ has no sense because
$y\in (a^{\circ}_j, b^{\circ}_j)$ and the images of all
$(a^{\circ}_j, b^{\circ}_j)$~s are greater than $c^{\circ}_1$. When
$r=s$ since $F(y)<d^{\circ}_s\leq F(\alpha^{\circ}_{s+1})=F(x_i)$,
the intermediate value theorem implies that there exists a point
$z\in (y,x_i]$ such that $F(z)=d^{\circ}_s$. But according to the
definition of $\beta^{\circ}_s$ we must have $z\leq
\beta^{\circ}_s$ which contradicts with the position of
$y$. Finally when $1\leq r\leq s-1$, since
$F(y)<d^{\circ}_r<F(\alpha^{\circ}_{r+1}) =c^{\circ}_{r+1}$, the
intermediate value theorem implies that there exists a point
$z'\in (y,\alpha^{\circ}_{r+1})$ such that $F(z')=d^{\circ}_r$ but
according to the definition of $\beta^{\circ}_r$ we must have
$z'\leq \beta^{\circ}_r$ which is a contradiction.

On the other hand in the second state we may have
$d^{\circ}_r<c^{\circ}_{r+1}<F(y)$ for $1\leq r\leq s-1$,
$c^{\circ}_1<F(y)$ for $r=0$ and $d^{\circ}_s<F(y)$ for $r=s$. The
case $r=s$ has no sense because the images of all $(a^{\circ}_j,
b^{\circ}_j)$~s are less than $d^{\circ}_s$. When $r=0$ since
$F(\beta^{\circ}_0)=F(x_{i-1})\leq c^{\circ}_1<F(y)$, the
intermediate value theorem implies that there exists a point $t\in
[x_{i-1},y)$ such that $F(t)=c^{\circ}_1$. But according to the
definition of $\alpha^{\circ}_1$ we must have
$\alpha^{\circ}_1\leq t$ which is in contradiction with the
position of $y$.

Finally when $1\leq r\leq s-1$, since
$F(\beta^{\circ}_r)=d^{\circ}_r<c^{\circ}_{r+1}<F(y)$, the
intermediate value theorem implies that there exists a point
$t'\in (\beta^{\circ}_r,y)$ such that $F(t')=c^{\circ}_{r+1}$ but
according to the definition of $\alpha^{\circ}_{r+1}$ we must have
$\alpha^{\circ}_{r+1}\leq t'$ which is a contradiction. Thus, our
claim is proved, that is, non of the points $a^{\circ}_j$ or
$b^{\circ}_j$ lie inside the intervals
$(\beta^{\circ}_r,\alpha^{\circ}_{r+1})$ or in another words, all
points $a^{\circ}_j$ and $b^{\circ}_j$ lie only inside intervals
$[\alpha^{\circ}_r,\beta^{\circ}_r]$.

The above fact admits the definition of partition $S_2$ as
\begin{eqnarray}\label{eq:20}
\hspace{1cm} S_2:~x_{i-1}&\!=\!&\beta^{\circ}_0\leq \alpha^{\circ}_1\leq a^{\circ}_1< b^{\circ}_1
<\cdots<a^{\circ}_{j_1}< b^{\circ}_{j_1}\leq \beta^{\circ}_1 \nonumber\\
&\!<\!& \alpha^{\circ}_2\leq
a^{\circ}_{j_1+1}<b^{\circ}_{j_1+1} <\cdots<a^{\circ}_{j_{2}}<
b^{\circ}_{j_{2}}\leq\beta^{\circ}_2 \nonumber\\
&\!<\!&\alpha^{\circ}_3 <\cdots
<\alpha^{\circ}_s\leq\cdots<a^{\circ}_m<b^{\circ}_m\leq\beta^{\circ}_s\leq
\alpha^{\circ}_{s+1}=x_i,
\end{eqnarray}
which is clearly a refinement of partition $S_1$ defined in
\eqref{eq:19}. Thus, according to our hypothesis we see that
$|F(S_2)|-|F(S_1)|<\epsilon$ which by a simple but careful
observation results in the following relation
\begin{align*}
\sum_{j=1}^m |F(b^{\circ}_j)-F(a^{\circ}_j)|<\epsilon +
\sum_{r=1}^s|F(\beta^{\circ}_r)-F(\alpha^{\circ}_r)|.
\end{align*}

Recalling the definitions of $\alpha^{\circ}_r$ and
$\beta^{\circ}_r$ and since $J^{\circ}_{k_r}=(c^{\circ}_r,
d^{\circ}_r)$, the above relation converts to
\begin{align*}
\sum_{j=1}^m |F(b^{\circ}_j)-F(a^{\circ}_j)|<\epsilon +
\sum_{r=1}^s\lambda(J^{\circ}_{k_r}),
\end{align*}
and due to relation \eqref{eq:10} one obtains
\begin{align*}
\sum_{j=1}^m |F(b^{\circ}_j)-F(a^{\circ}_j)|<2\,\epsilon.
\end{align*}

Since the above relation is true for the end points of any finite
number (here $m$) of $I^{\circ}_{kl}$~s, it is also valid for all
of them, that is
\begin{eqnarray} \label{eq:21}
\sum_{k,l} |F(b^{\circ}_{kl})-F(a^{\circ}_{kl})|<2\,\epsilon.
\end{eqnarray}

Now by gathering the relations \eqref{eq:15}, \eqref{eq:16} and
\eqref{eq:21} it is found that
\begin{eqnarray}\label{eq:22}
\sum^{\infty}_{k,l=1} |F(b_{kl})-F(a_{kl})|<4\,\epsilon.
\end{eqnarray}

Inequalities \eqref{eq:11}, \eqref{eq:13} and
\eqref{eq:22} yield
\begin{eqnarray}\label{eq:23}
\lambda(p(N_i))\leq\sum^{\infty}_{k,l=1}
\lambda(p(I_{kl}))<5\,\epsilon.
\end{eqnarray}

This establishes the zero measure of $p(N_i)$ when $F(x_{i-1})\neq
F(x_i)$.

It only remains to show for the non-decreasing function $n=p-F$,
that $\lambda(n(N_i))=0$. In an exactly similar way of obtaining
relation \eqref{eq:11} one easily finds that
\begin{align*}
\lambda(n(N_i))\leq \sum_{k,l=1}^{\infty} \lambda(n(I_{kl})),
\end{align*}
where still $I_{kl}=(a_{kl},b_{kl})$ and thus
$n(I_{kl})\subset[n(a_{kl}),n(b_{kl})]$. Then we notice that for
any two points $x,y\in[x_{i-1},x_i]$ since $n=p-F$ we have
\begin{align*}
|n(y)-n(x)|\leq |p(y)-p(x)|+|F(y)-F(x)|.
\end{align*}

Substituting $a_{kl},b_{kl}$~s to resp.
$x,y$ the latter relation yields
\begin{align*}
\lambda(n(N_i))\leq\sum_{k,l=1}^{\infty} \lambda(n(I_{kl}))\leq
\sum_{k,l=1}^{\infty} \lambda(p(I_{kl}))+\sum^{\infty}_{k,l=1}
|F(b_{kl})-F(a_{kl})|.
\end{align*}

The upper bounds for the first and second terms on the right hand
side of the above relation due to \eqref{eq:22} and \eqref{eq:23}
proves the zero measure of $n(N_i)$.


\section{\bf Conclusion}
\vskip 0.4 true cm

As one little step towards understanding the regularity of
hydrodynamical quantities, it was attempted to see a more direct
and clear dependence of continuity and integrability through the
Lebesgue integral while there is a hope to generalize the method
to find the situation for other types of integration. Indeed,
there probably exists an alternative kind of absolute continuity
in connection with other types of integration rather than the
Lebesgue one.

Even further, since the used method here essentially employed the
general measure-type informations, it looks to have sense to
include the issue of measurability of fluid functions under the
mechanism of singularity. In other words, the problem of blow up
usually deals with singularities and therefore infinite integrals
while it is not yet known if this dynamics can change even the
measurability of solutions or not.

It was seen here that the absolute continuity can be extracted
directly as a consequence of measure-type properties of functions.
There was nowhere used the idea of differentiability which is the
result of the Vitali covering lemma. Instead, the Radon--Nikodym
theorem was the main tool which relies solely on the excellent
consistency between the Lebesgue integral and a measure space.

In addition, Lemma \ref{4} was proven showing a slightly more general
result than needed for the proof of the Banach-Zarecki theorem.
Although the classical version of this theorem was proven here but
it is not surprising if one can generalize this proof to more
general spaces and even higher dimensions.


\vskip 0.4 true cm

\begin{center}{\textbf{Acknowledgments}}
\end{center}
The authors wish to sincerely appreciate the referee for useful comments and suggestions 
which definitely improved the paper.  \\ \\
\vskip 0.4 true cm




\bigskip
\bigskip


{\footnotesize \pn{\bf A. Mahdipour-Shirayeh}\; \\ {Department of
Mathematics}, {Iran University
of Science and Technology,\\ Tehran 16846-13114, Iran}\\
{\tt Email: mahdipour@iust.ac.ir}\\

{\footnotesize \pn{\bf H. Eshraghi}\; \\ {Department of
Physics}, {Iran University
of Science and Technology,\\ Tehran 16846-13114, Iran}\\
{\tt Email: eshraghi@iust.ac.ir}\\
\end{document}